# Estimation of material parameter uncertainties using probabilistic and interval approaches

Thomas Most (Bauhaus-Universität Weimar, Germany)

## 1 Introduction

Within the calibration of material models, often the numerical results of a simulation model **y** are compared with the experimental measurements **y***. Usually, the differences between measurements and simulation are minimized using least squares approaches including global and local optimization techniques. In this paper, the resulting scatter or uncertainty of the identified material parameters **p** are investigated by assuming the measurement curves as non-deterministic. Based on classical probabilistic approaches as the Markov estimator or the Bayesian updating procedure, the scatter of the identified parameters can be estimated as a multi-variate probability density function. Both procedures require a sufficient accurate knowledge or estimate of the scatter of the measurement points, often modeled by a Gaussian covariance matrix.

In this paper, we present a different idea by assuming the scatter of the measurements not as correlated random numbers but just as individual intervals with known minimum and maximum values. The corresponding possible minimum and maximum values for each input parameter can be obtained for this assumption using a constrained optimization approach. However, the identification of the whole possible parameter domain for the given measurement bounds is not straight forward. Therefore, we introduce an efficient line-search method, where not only the bounds itself but also the shape of the feasible parameter domain can be identified. This enables the quantification of the interaction and the uniqueness of the input parameters. As a numerical example, we identify the fracture parameters of plain concrete with respect to a wedge splitting test, where five unknown material parameters could be identified.

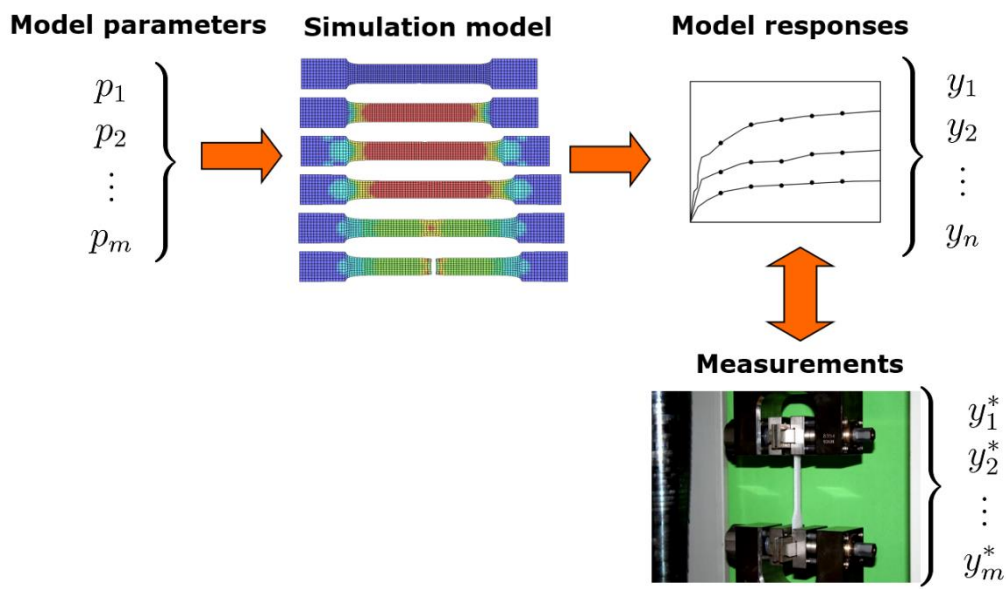

Fig. 1: Comparison of the results of a simulation model **y** by using a given parameter set **p** with measurement observation **y***

## 2 Probabilistic approaches

### 2.1 Maximum likelihood approach

In the calibration task, usually a simulation model is used to estimate unknown parameters by minimizing the deviation between the model responses and measurement observations. In the maximum likelihood approach [1,2] it is assumed that the simulation model can represent the physical phenomena perfectly and thus the resulting difference between measurements and simulation response are caused only by measurement uncertainties. Generally, a joint normal distribution is assumed for the measurement scatter

$$P(\mathbf{y}^* - \mathbf{y}) = \frac{1}{\sqrt{(2\pi)^m |\mathbf{C_{yy}}|}} \exp\left[-\frac{1}{2}(\mathbf{y}^* - \mathbf{y})^T \mathbf{C_{yy}}^{-1}(\mathbf{y}^* - \mathbf{y})\right]$$

where $\mathbf{C_{yy}}$ is the covariance matrix of the measurement uncertainty. If the likelihood of the parameters is maximized, we obtained the following objective function, which has to be minimized

$$J = (\mathbf{y}^* - \mathbf{y})^T \mathbf{C_{yy}}^{-1} (\mathbf{y}^* - \mathbf{y}) \to \min$$

By linearizing this objective, we obtain the following updating scheme

$$\Delta \mathbf{p} = \left(\mathbf{A}^T \mathbf{C_{yy}}^{-1} \mathbf{A}\right)^{-1} \mathbf{A}^T \mathbf{C_{yy}}^{-1} \Delta \mathbf{y}, \quad \mathbf{A} = \frac{\partial \mathbf{y}}{\partial \mathbf{p}}$$

where **A** is the local sensitivity matrix containing the derivatives of the individual response values **y** w.r.t. the model parameters **p**.

### 2.2 Markov estimator

Assuming, that the optimal parameter set $\mathbf{p_{opt}}$ has been found by e.g. a combination of a global and local optimization approaches, we can calculate the sensitivity matrix at the optimal parameter set and estimate the covariance matrix of the parameters [2]

$$\mathbf{C_{pp}} = \left(\mathbf{A}_{opt}^T \mathbf{C_{yy}}^{-1} \mathbf{A}_{opt}\right)^{-1}$$

This covariance matrix $\mathbf{C_{pp}}$ contains the necessary information about the scatter of the individual parameters as well as possible pairwise correlations. The distribution of the parameters is normal as assumed for the measurements. The estimated scatter and correlations are a very useful information to judge about the accuracy and uniqueness of the identified parameters in a calibration process. This method can be interpreted as an inverse first order second moment method: assuming a linear relation between parameters and measurements and a joint normal distribution of the measurements, we can directly estimate the joint normal covariance of the parameters. However, the application of this linearization scheme requires an accurate estimate of the optimal parameter set by a previous least squares minimization.

### 2.3 Bayesian updating

Another probabilistic method is the Bayesian updating approach [1], which considers nonlinear relations between the parameters and the responses but requires much more numerical effort to obtain the statistical estimates. In this approach the unknown conditional distribution of the parameters w.r.t. to the given measurements is formulated using the Bayes' theorem as

$$P(\mathbf{p}|\mathbf{y}^*) = \frac{P(\mathbf{y}^*|\mathbf{p}) \cdot P(\mathbf{p})}{P(\mathbf{y}^*)}$$

where the likelihood function can be formulated as multi-normal probability density function similar to the Markov estimator

$$P(\mathbf{y}^*|\mathbf{p}) = \frac{1}{\sqrt{(2\pi)^m |\mathbf{C_{yy}}|}} \exp\left[-\frac{1}{2}(\mathbf{y}^* - \mathbf{y})^T \mathbf{C_{yy}}^{-1}(\mathbf{y}^* - \mathbf{y})\right]$$

Since the normalization term *P*(**y**\*) is usually not known, the posterior parameter distribution *P*(**p**|**y**\*) is obtained by sampling procedures such as the Metropolis-Hastings algorithm [3]. The prior distribution

*P*(**p**) can be assumed either from previous information or as uniform distribution within assumed parameter bounds. Further details on this procedure can be found e.g. in [4].

The challenge in both probabilistic approaches is the estimate of the covariance matrix of the measurement uncertainty **C**$_{yy}$. If a measurement procedure is repeated very often, this matrix can be estimated directly from measurement statistics. However, this is often not possible. A valid estimate could be to assume the measurement errors to be independent. The resulting covariance matrix is then diagonal, where at least the scatter of the individual measurement points has to be estimated. From a small number of measurement repetitions this can be estimated straight forward. If even such information is not available, the measurement error can be assumed to be constant for each output value

$$\mathbf{C_{yy}} = \sigma_y^2 \mathbf{I} \qquad \mathbf{C_{pp}} = \sigma_y^2 \left(\mathbf{A}_{opt}^T \mathbf{A}_{opt}\right)^{-1}$$

In this case, the estimated covariances of the parameters is directly scaled by the assumed measurement scatter. In case, that only a single measurement curve is available for the investigated responses, the misfit of the calibration process can be used for a rough estimate. Assuming the measurement errors to be constant and independent, we obtained the well-known least squares formulation.

## 3    Interval line-search approach

Instead of modeling the measurement points as correlated random numbers, which requires the estimate of the full covariance matrix, we can assume, that each measurement output is an interval number with known minimum and maximum bounds as shown in figure 2.

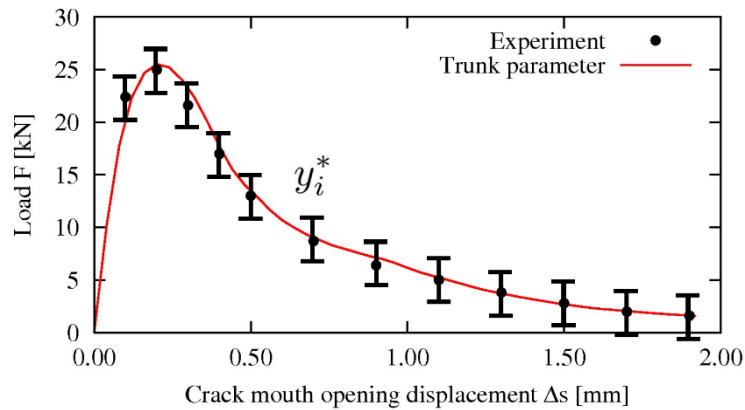

Fig. 2: Assuming the uncertain measurement data as interval numbers for the identification of the input parameter bounds

In the classical interval optimization approach, usually the interval of an uncertain response is obtained from the known interval of the input parameters by several forward optimization steps. This is the so-called α-level optimization [6] and is shown in principle in figure 3: the most probable value is assumed for each input parameter with the maximum α-level and a linear descent to zero at the lower and upper bounds. The α-level optimization will then detect the shape of the corresponding α-levels of the unknown outputs. In the inverse approach, we now search for the possible minimum and maximum values of the input parameters **p** within the given level, while each of the response value **y** is constrained to the corresponding interval. This can be solved for each input parameter and each α-level individually by a constrained optimization run with all unknown input parameters as optimization design variables.

This optimization, which searches for the bounds of each input parameter separately cannot investigate the interactions or correlations between the possible input parameter combinations. Therefore, we suggest a new interval-search approach, which investigates the whole joint domain of the input parameters. Based on the optimal parameter set **p**$_{opt}$, found by a previous deterministic calibration, the input parameter space is discretized by a given number of uniformly distributed circular search directions as shown in fig 4. For each of these directions a line-search is applied to find the boundary of the feasible parameter domain w.r.t. to the given interval bounds of the measurement points. This approach is similar to the directional sampling method known from the reliability analysis [7,8].

In order to extend this method to more than two input parameters, the search directions have to be discretized on a 3D- or higher dimensional (hyper-) sphere. One very efficient discretization method to generate such direction vectors on the hyper-sphere are the so-called Fekete points [8], which are generated using a minimum potential energy criterion. This results in very equally distributed points as shown in fig 4. This optimized discretization will require a significant smaller number of model evaluations to obtain a suitable representation of the parameter domain boundary. For each of the search directions then a line-search is performed to obtain the boundary of the feasible parameter domain. Since this is an independent one-dimensional search for every direction point on the hyper-sphere is can be efficiently computed. The final points on the boundary can be used to obtain a discretized shape of the feasible parameter domain. If more than two or three parameters are investigated, the visualization of the parameter domain can only be displayed in a subspace.

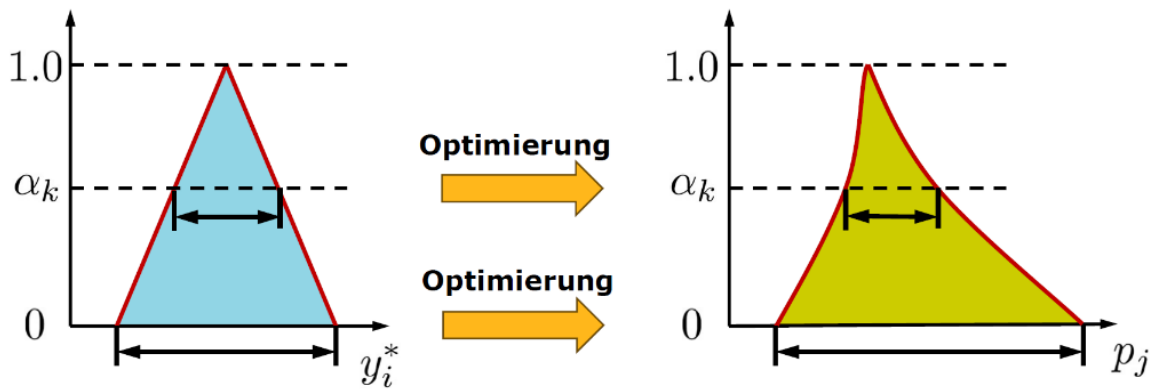

Fig. 3: Interval optimization applied to the inverse identification of minimum and maximum parameter bounds based on given measurement intervals

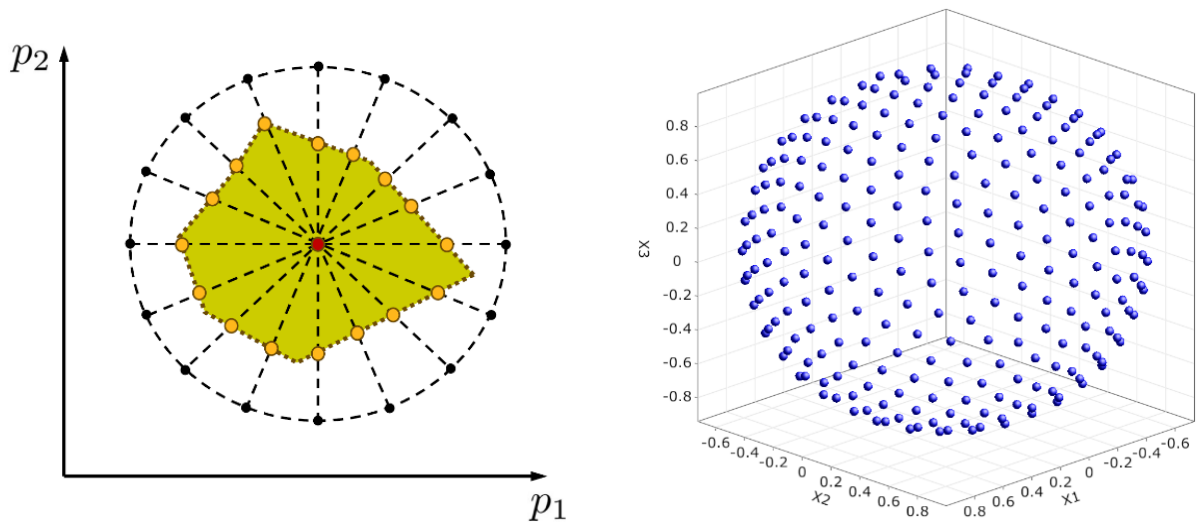

Fig. 4: Identification of the feasible input parameter domain using the interval line-search approach in two dimensions (left) and using Fekete points on the hyper-sphere in higher dimensions (right)

## 4 Numerical example

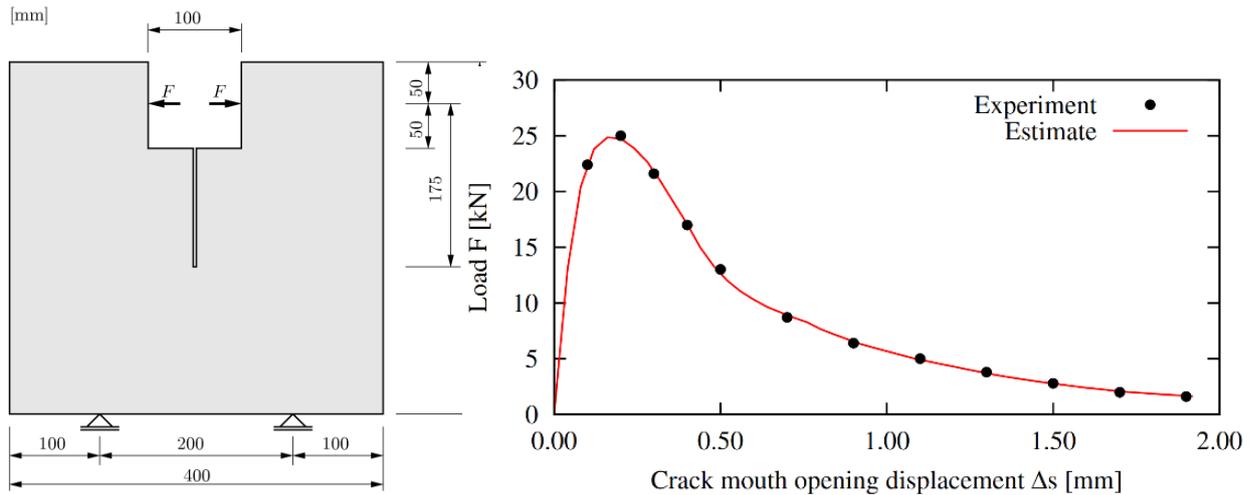

Fig. 5: Investigated wedge-splitting test setup according to [9] (left), comparison of measurements and calibrated simulation model (right)

By means of this numerical example the material parameter of plain concrete are identified using the measurement results of a wedge splitting test published in [9]. Six unknown parameters are investigated as the Young's modulus *Emod*, the Poisson's ratio *nue*, the tensile strength *fct*, the Mode-I fracture energy *Gf* and the two shape parameters *alpha_ft* und *alpha_wc* of a bilinear softening law. The geometry of the investigated concrete specimen is shown in figure 5. The structure is discretized by 2D finite elements and a predefined straight crack and the softening curve is obtained by a displacement-controlled simulation. Further details about the simulation model can be found in [10]. In a first step using a forward sensitivity analysis [5] it was observed, that the Poisson's ratio does not have any influence on the simulation curve. Therefore, it cannot be identified with the given experimental setup and is not considered in the following investigations.

In a second step the remaining five parameters are calibrated with a global and local optimization run using a least squares formulation with constant measurement errors for the objective function. The simulation curve of the optimal parameter set is compared to the measurement points in figure 5. The corresponding Root-Mean-Square-Error (RMSE) is 151 N, which is approximately 0.6% of the maximum force.

In a third step this RMSE is used to estimate a diagonal measurement covariance matrix as the standard deviation for all measurement points. Based on the covariance matrix of the measurements, the covariance matrix of the input parameters is estimated by the Markov estimator assuming a normal distribution for all parameters. In figure 6 the generated samples of the multi-variate Gaussian probability density are shown for two subspaces. The figure indicates, that a significant correlation between the identified Young's modulus and the tensile strength can be observed, meanwhile the fracture energy is almost uncorrelated w.r.t. the tensile strength. Additionally, to the Markov estimator, we apply the Bayesian updating procedure using the same measurement covariance matrix. In figure 6 the samples obtained by the Metropolis-Hastings algorithm are shown. The figure indicates, that the distribution of the Young's modulus and the tensile strength is not symmetric as assumed by the Markov estimator.

Finally, the introduced interval search algorithm is applied in this example. In a first step only the Young's modulus and the tensile strength are considered as uncertain parameters meanwhile the remaining parameters are kept constant. The resulting two-dimensional search points are shown for different sizes of the measurement intervals in figure 7. The figure clearly indicates, that the obtained parameter domain boundary is not symmetric as assumed by the Markov estimator. The assumed interval size of +/- 500 N corresponds 3.3 times the RMSE, which is equivalent to the 99.9% confidence interval. In a second step all five parameters are considered in the interval search. The identified points on the boundary are shown additionally in figure 7 in the same subspace as used in the previous investigation. The identified boundary of the input parameters in the five-dimensional case is slightly extended for small values of the Young's modulus compared to the 2D analysis.

As a result of the presented investigations, we can summarize, that all approaches provide in general similar results. The Markov estimator is most efficient but assumes normally distributed input parameters. The presented interval search approach does not require any knowledge on the measurement covariance matrix but requires much more model evaluations similar as the Bayesian updating approach.

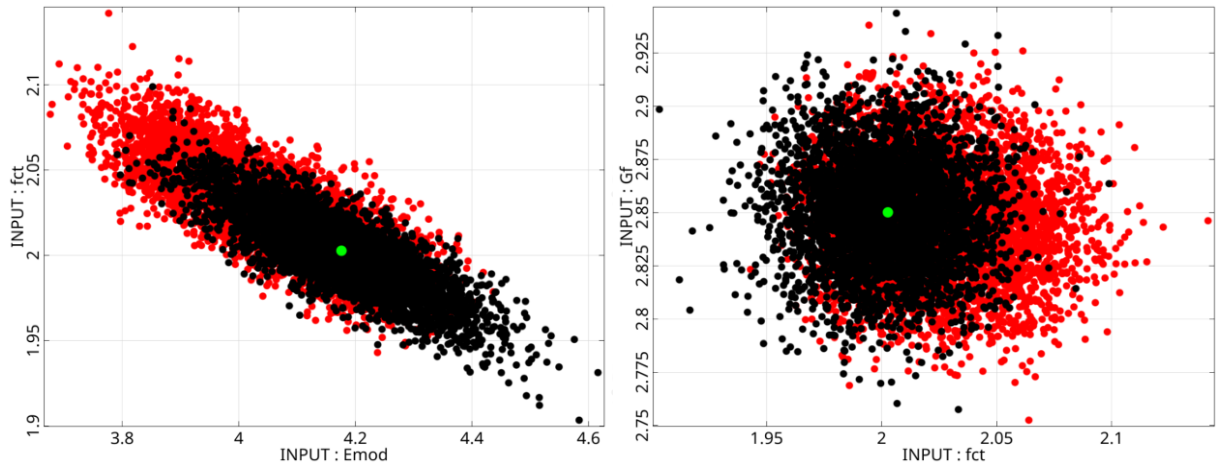

Fig. 6: Estimated parameter scatter and correlations for the Young's modulus (Emod), the tensile strength (fct) and the specific fracture energy (Gf) using the Markov estimator (black) and the Bayesian updating approach (red)

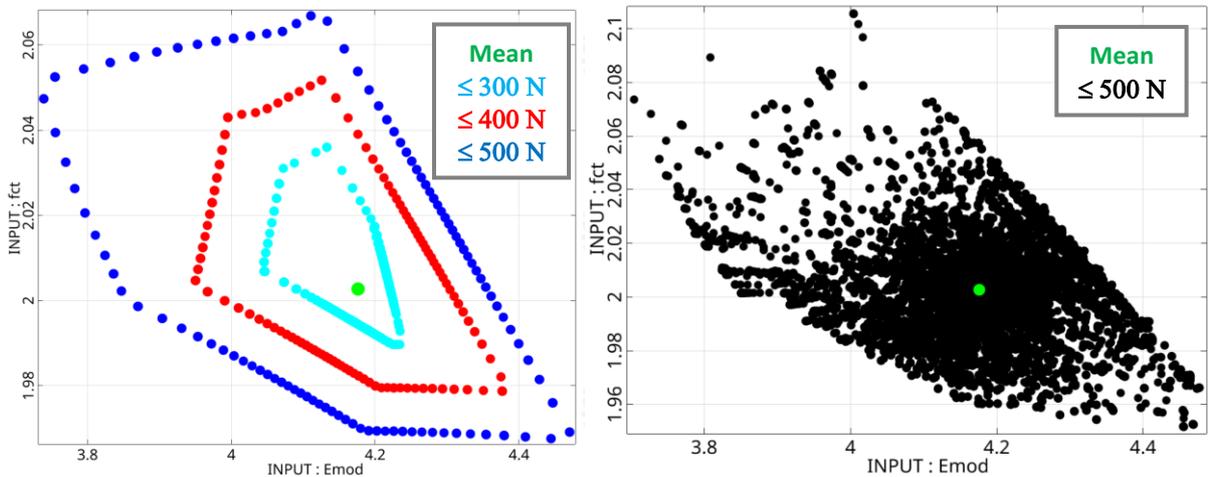

Fig. 7: Estimated α-level bounds in the subspace of the Young's modulus and the tensile strength considering only the two parameters in the line-search (left) and considering all five parameters (right)

# 5   References

[1]   Beck, J. V. and K. Arnold (1977). *Parameter estimation in engineering and science.* New York: Wiley Interscience
[2]   Ledesma, A., A. Gens, and E. E. Alonso (1996). *Estimation of parameters in geotechnical backanalysis – I. Maximum likelihood approach.* Computers and Geotechnics 18, 1–27.
[3]   Hastings, W. (1970). *Monte Carlo sampling methods using Markov chains and their applications.* Biometrika 57, 97–109.
[4]   Most, T. (2010). *Identification of the parameters of complex constitutive models: Least squares minimization vs. Bayesian updating.* IFIP Working Group Conference, München, 2010
[5]   Most, T.; Kallmeyer, R.; Niemeier, R. (2019). *Estimate of Material Parameter Uncertainties in calibrated Simulation Models.*, NAFEMS World Congress, Quebeq, Kanada, 2019
[6]   Möller, B.; Beer, M. (2000). *Fuzzy structural analysis using α-level optimization.* Computational Mechanics, 26:547-565
[7]   Bucher, C. (2009). *Computational analysis of randomness in structural mechanics.* Structures and infrastructures book series, Vol. 3. CRC Press.
[8]   Nie, J.; Ellingwood, B.R. (2000). *Directional methods for structural reliability analysis.* Structural Safety 22, 233-249
[9]   Trunk, B. (1999). *Einfluss der Bauteilgrösse auf die Bruchenergie von Beton.* Dissertation, Eidgenössische Technische Hochschule, Zürich
[10]  Most, T. (2005). *Stochastic crack growth simulation in reinforced concrete structures by means of coupled finite element and meshless methods,* Dissertation, Bauhaus-Universität Weimar